
\magnification=1200
\hoffset=.0cm
\voffset=.0cm
\baselineskip=.55cm plus .55mm minus .55mm

%
%
\def\ref#1{\lbrack#1\rbrack}
%
%
%
%
\input amssym.def
\input amssym.tex
%
%
\font\teneusm=eusm10
\font\seveneusm=eusm7
\font\fiveeusm=eusm5
%
%
\font\sf=cmss10

%
%
\newfam\eusmfam
\textfont\eusmfam=\teneusm
\scriptfont\eusmfam=\seveneusm
\scriptscriptfont\eusmfam=\fiveeusm
\def\sans#1{\hbox{\sf #1}}

%
%
%
%

\def\tr{{\rm tr}\hskip 1pt}

\def\ad{{\rm ad}\hskip 1pt}
\def\id{{\rm id}\hskip 1pt}
\def\ker{{\rm ker}\hskip 1pt}
\def\Ad{{\rm Ad}\hskip 1pt}

\def\Gau{{\rm Gau}\hskip 1pt}
\def\Conf{{\rm Conf}\hskip 1pt}
\def\Lie{{\rm Lie}\hskip 1pt}
%
%
%

\magnification=1200
\hoffset=-.6cm
\baselineskip= .55cm plus .055cm minus .055cm
\vsize=23.8 truecm
\hsize=16.5 truecm
\nopagenumbers

\hbox to 16.5 truecm{October 1993   \hfil DFUB--93--13}
\hbox to 16.5 truecm{Version 1  \hfil}
\vskip2cm
\centerline{{\bf A KRICHEVER--NOVIKOV FORMULATION OF CLASSICAL}}
\centerline{$W$ {\bf ALGEBRAS ON RIEMANN SURFACES}}
\vskip1cm
\centerline{by}
\vskip.5cm
\centerline{\bf Roberto Zucchini}
\centerline{\it Dipartimento di Fisica, Universit\`a degli Studi di Bologna}
\centerline{\it V. Irnerio 46, I-40126 Bologna, Italy}
\vskip1cm
\centerline{\bf Abstract}
\vskip.4cm
\noindent
It is shown how the theory of classical $W$--algebras can be formulated on
a higher genus Riemann surface in the spirit of Krichever and Novikov.
An intriguing relation between the theory of $A_1$ embeddings into
simple Lie algebras and the holomorphic geometry of Riemann surfaces is
exihibited.
\vfill\eject
\vskip.6cm
\item{0.} {\bf Introduction}
\vskip.4cm
\par
Recently, a large body of literature has been devoted to the development
of the theory of $W$--algebras (see refs. \ref{1,2,3} for a
comprehensive review). Such studies show that one can associate a
$W$--algebra ${\cal W}^{\goth g}_{\goth t}$ to any non trivial $A_1$
subalgebra $\goth t$ of the Lie algebra $\goth g$ of a
complex simple Lie group $G$. The reduction
of the adjoint representation of $\goth g$ with respect to $\goth t$ plays
a fundamental role. The issue of the potentially non trivial relation between
the algebraic structure of ${\cal W}^{\goth g}_{\goth t}$ and the geometry
and the topology of the base surface is not addressed since the latter is
assumed to be merely a sphere with two punctures. A few years ago,
I. M. Krichever and S. P. Novikov showed that there exist generalizations of
the Heisenberg and Virasoro algebra on any compact connected oriented
Riemann surface of genus $\ell\geq 2$ with two distinguished points
$P_+$ and $P_-$ in general position \ref{4,5} (see also ref. \ref{6} for a
pedagogical introduction). Their construction was subsequently generalized
to super Virasoro \ref{7,8} and Kac--Moody \ref{9} algebras.
In this communication, I shall outline
a formulation \`a la Krichever--Novikov (KN) of classical $W$--algebras.
An intriguing relation between the theory of the $A_1$ embeddings into
simple Lie algebras and the holomorphic geometry of Riemann surfaces
will emerge (see ref. \ref{10} for an early attempt in this direction).
The theory of reduction of Poisson manifolds is an essential
ingredient.
\par\vskip.6cm
\item{1.} {\bf The spaces} ${\sans K\sans N}_j$ {\bf and} ${\sans W}_j$
{\bf and the Drinfeld--Sokolov holomorphic vector bundle}
\vskip.4cm
\par
Let us briefly recall the basic notions of the KN theory
\footnote{${}^1$}{The conventions adopted in this paper are the following.
I denote by ${\cal O}(K)$ the sheaf of germs of holomorphic sections
of a $1$ cocycle $K$ on $\Sigma$ and by $\Gamma(U,{\cal O}(K))$ the space of
sections of ${\cal O}(K)$ on a open set $U$ of $\Sigma$. I denote by
$k$ the holomorphic canonical line bundle of $\Sigma$.
For any $(1,0)$ connection $C$ of $K$, $C_b=k_{ba}(\Ad K_{ba}C_a
+\partial_bK_{ba}K_{ba}{}^{-1})$, where $a,b,c,\ldots$ are patch indices.
For any projective connection $R$, $R_b=k_{ba}{}^2(R_a-\{z_b,z_a\})$.}.
Let $\Sigma^\circ=\Sigma\setminus\{P_+,P_-\}$. The basic functional spaces
of the KN theory are the KN spaces
$${\sans K\sans N}_j=\Gamma(\Sigma^\circ,{\cal O}(k^{\otimes j})),\eqno(1.1)$$
where $j\in{\Bbb Z}/2$.
There is a non singular pairing of
${\sans K\sans N}_j$ and ${\sans K\sans N}_{1-j}$, the KN pairing, defined by
$$\langle p,q\rangle=\oint_{C_\tau}{dz\over 2\pi i}pq,\quad
p\in{\sans K\sans N}_j,~q\in{\sans K\sans N}_{1-j},\eqno(1.2)$$
where $C_\tau$ is the curve in $\Sigma$ of KN time $\tau\in{\Bbb R}$. The
pairing is actually independent from $\tau$, since the curves $C_\tau$
for varying $\tau$ are all homologous \ref{4,5}. For each $j$, the space
${\sans K\sans N}_j$ admits a standard basis, the KN basis, which is very
useful in calculations (see ref. \ref{6} for explicit expressions in terms of
theta functions). The KN bases of ${\sans K\sans N}_j$ and ${\sans K
\sans N}_{1-j}$ are dual with respect to the pairing $(1.2)$.

The basic symmetry group of the KN theory is the group $\Conf_0(\Sigma^\circ)$
of holomorphic diffeomorphisms $f$ of $\Sigma^\circ$ onto itself
continuously connected to $\id_\Sigma$. Its Lie algebra is
$\Lie\Conf_0(\Sigma^\circ)={\sans K\sans N}_{-1}$. The Lie brackets are given
by
$$[v,w]=v\partial w-w\partial v,\quad v,w\in\Lie\Conf_0(\Sigma^\circ).
\eqno(1.3)$$
$\Conf_0(\Sigma^\circ)$ acts on the KN spaces ${\sans K\sans N}_j$
$$f^*p=(\partial f)^jp\circ f,\quad
f\in\Conf_0(\Sigma^\circ),~p\in{\sans K\sans N}_j.\eqno(1.4)$$
At the infinitesimal level, this relation reduces into
$$\theta_vp=v\partial p+j(\partial v)p,\quad
v\in\Lie\Conf_0(\Sigma^\circ),\quad p\in{\sans K\sans N}_j.\eqno(1.5)$$
The KN pairing $(1.2)$ is invariant under $\Conf_0(\Sigma^\circ)$.

The above framework can be extended as follows. Let $t_{-1},t_0,t_{+1}$
be the standard generators of a $A_1$ subalgebra $\goth t$ of the complex
simple Lie algebra $\goth g$. They satisfy the relations
$$[t_{+1},t_{-1}]=2t_0,\quad [t_0, t_{\pm 1}]=\pm t_{\pm 1}.\eqno(1.6)$$
The adjoint representation of $\goth g$ is completely reducible with
respect to the subalgebra $\goth t$. Let us denote by $\Pi$ the set of the
representations of $A_1$ appearing in the reduction, each counted with its
multiplicity, by $j_\eta$ the spin of a representation $\eta\in\Pi$
and by $I_\eta$ the set $\{m|m\in{\Bbb Z}/2,|m|\leq j_\eta,
m-j_\eta\in{\Bbb Z}\}$. To a
representation $\eta\in\Pi$, there is associated a
distinguished set of generators $t_{\eta,m}$, $m\in I_\eta$
of $\goth g$ such that
$$[t_d,t_{\eta,m}]=C^d{}_{j_\eta,m}t_{\eta,m+d},
\quad d=-1,0,+1,\eqno(1.7a)$$
$$C^{\pm 1}{}_{j,m}=[j(j+1)-m(m\pm 1)]^{1\over 2},\quad C^0{}_{j,m}=m.
\eqno(1.7b)$$
$t_{-1},t_0,t_{+1}$
themselves span a representation $o\in\Pi$ with $j_o=1$ and
$t_{o,\pm 1}=\mp 2^{-{1\over 2}}t_{\pm 1}$ and $t_{o,0}=t_0$.
The non degeneracy of the Cartan form of $\goth g$ implies that
each representation $\eta\in\Pi$ admits a conjugate representation
$\bar\eta$. $\bar{\bar\eta}=\eta$, $j_\eta=j_{\bar\eta}$ and
$\bar\eta=\eta$ if and only if $j_\eta\in{\Bbb Z}$.
For any representation $D$ of $\goth g$,
$$\tr_D\big(t_{\eta,m}t_{\zeta,-n}\big)=N_{D\eta}
(-1)^{j_\eta-m}\delta_{\bar\eta,\zeta}\delta_{m,n},\eqno(1.8)$$
where $N_{D\eta}$ is a normalization constant such that
$N_{D\bar\eta}=(-1)^{2j_\eta}N_{D\eta}$. In what follows, I shall
assume tacitly that all Lie algebra and group elements are taken in the
fundamental defining representation of $\goth g$.

The Drinfeld--Sokolov vector bundle is defined by
$$L_{ab}=k_{ab}{}^{-t_0}
\exp(\partial_ak_{ab}{}^{-1}t_{-1})\eqno(1.9)$$
\ref{11}. $L$ possesses a distinguished  $(1,0)$ holomorphic connection,
the Drinfeld--Sokolov connection, given by
$$A=(1/2)t_{+1}-Rt_{-1},\eqno(1.10)$$
where $R$ is a reference holomorphic projective connection
\ref{11}. The structure of this connection justifies the name adopted
for $L$ and $A$ \ref{12}. The existence of $A$ shows the flatness
of $L$. The flat structures of $L$ are parametrized by the
$(1,0)$ holomorphic connections $A_{\zeta}$ of the form
$A(\zeta)=A+\sum_{\eta\in\Pi}\zeta_\eta t_{\eta,-j_\eta}$,
where $\zeta_\eta\in\Gamma(\Sigma,{\cal O}(k^{\otimes j_\eta+1}))$
\ref{13}. $L$ is unstable, since $\sum_{\eta\in\Pi}\zeta_\eta t_{\eta,-j_\eta}$
is a non trivial element of $\Gamma(\Sigma,{\cal O}(\Ad L))$ \ref{13}.

In the present context, besides the KN spaces, one needs the spaces
$${\sans W}_j=\Gamma(\Sigma^\circ,{\cal O}(k^{\otimes j}\otimes \Ad L)),
\eqno(1.11)$$
where $j\in{\Bbb Z}/2$. There is a non singular pairing of ${\sans W}_j$
and ${\sans W}_{1-j}$ given by
$$\langle X,W\rangle=\oint_{C_\tau}{dz\over 2\pi i}\tr(XW),\quad
X\in{\sans W}_j,~W\in{\sans W}_{1-j}.\eqno(1.12)$$
The spaces ${\sans W}_j$ admit standard bases. However, since here
one aims at classical $W$--algebras, such bases are not necessary
and the KN basis of the KN spaces ${\sans K\sans N}_j$ suffice
for calculations.

Denote by $\Gau_0(\Sigma^\circ,L)$ the group of $G$--valued
gauge transformations
$\gamma$ of $L$ holomorphic off $P_+$ and $P_-$ and continuously
connected to the identity. Its Lie algebra $\Lie\Gau_0(\Sigma^\circ,L)$
is ${\sans W}_0$ with Lie brackets
$$[\xi,\eta]=[e(\xi),e(\eta)],\quad \xi,\eta\in \Lie\Gau_0(\Sigma^\circ,L),
\eqno(1.13)$$
where in the right hand side $e$ is the evaluation map at a given point of
$\Sigma^\circ$ and the Lie brackets are those of $\goth g$.
$\Gau_0(\Sigma^\circ,L)$ acts on ${\sans W}_j$ through the adjoint
representation
$$\gamma W=\Ad\gamma W, \quad\gamma\in\Gau_0(\Sigma^\circ,L),
{}~W\in{\sans W}_j.\eqno(1.14)$$
At the infinitesimal level, this relation becomes
$$\delta_\xi W=[\xi,W], \quad\xi\in\Lie\Gau_0(\Sigma^\circ,L),
{}~W\in{\sans W}_j.\eqno(1.15)$$
The pairing $(1.12)$ is invariant under $\Gau_0(\Sigma^\circ,L)$.

The action of $\Conf_0(\Sigma^\circ)$ extends naturally to the spaces
${\sans W}_j$:
$$f^*W=(\partial f)^j\Ad L_f W\circ f,\quad f\in\Conf_0(\Sigma^\circ),
{}~W\in{\sans W}_j,\eqno(1.16a)$$
$$L_f=(\partial f)^{-t_0}
\exp(\partial(\partial f)^{-1}t_{-1}).\eqno(1.16b)$$
At the infinitesimal level, this relation reduces into
$$\theta_vW=v\partial_AW+j(\partial v)W+[L_v,W],
\quad v\in\Lie\Conf_0(\Sigma^\circ),~W\in{\sans W}_j,\eqno(1.17a)$$
$$L_v={\cal D}v,\quad{\cal D}=(1/2)t_{+1}-\partial t_0-(\partial^2+R)t_{-1},
\eqno(1.17b)$$
where $\partial_A=\partial-\ad A$ is the covariant derivative of $A$.
One can show that $L_v\in{\sans W}_0$ and that $L_v$ satisfies the
important equation
$$\partial_AL_v=-D_1vt_{-1},\quad D_1=\partial^3+2R\partial+(\partial R),
\eqno(1.18)$$
where $D_1$ is a Bol operator \ref{14}. The pairing $(1.12)$ is invariant
under $\Conf_0(\Sigma^\circ)$.
\par\vskip.6cm
\item{2.} {\bf The Poisson manifold} $({\sans W}_1,\{\cdot,\cdot\}_\kappa)$
\vskip.4cm
\par
${\sans W}_1$ can be endowed with a Poisson structure depending on a
parameter $\kappa\in{\Bbb C}$. The Poisson structure is completely defined by
giving the Poisson brackets of the linear functionals on ${\sans W}_1$.
The Poisson brackets of general functions on ${\sans W}_1$ are obtained
by enforcing the Leibniz rule. Since the pairing $(1.12)$ is non singular,
every linear functional on ${\sans W}_1$ is of the form
$$\lambda_X(W)=\langle X,W\rangle,\quad W\in{\sans W}_1,\eqno(2.1)$$
for some $X\in{\sans W}_0$. One sets
$$\{\lambda_X+a,\lambda_Y+b\}_\kappa=\lambda_{[X,Y]}+\kappa\chi(X,Y),
\quad X,Y\in{\sans W}_0,~a,b\in{\Bbb C}\eqno(2.2a)$$
$$\chi(X,Y)=\langle X,\partial_AY\rangle.\eqno(2.2b)$$
It is straightforward to verify that the Poisson brackets
$\{\cdot,\cdot\}_\kappa$ are bilinear, antisymmetric and satisfy the Jacobi
identity as the should. The above Poisson structure clearly resembles that
of Kac--Moody phase space. Namely, the level and the Kac--Moody current would
be $-\kappa$ and $\kappa A+W$, respectively. However, the geometrical
interpretation is completely different as the relevance of the
Drinfeld--Sokolov vector bundle shows.

For any $\gamma\in\Gau_0(\Sigma^\circ,L)$, one has that $\chi(\gamma X,
\gamma Y)=\chi(X,Y)-\langle [X,Y],\gamma^{-1}\partial_A\gamma\rangle$.
So, the ordinary action of $\Gau_0(\Sigma^\circ,L)$ on ${\sans W}_1$, defined
by $(1.14)$, is not Poisson: it does not leave the Poisson brackets invariant.
However, there exists a deformation of the action enjoying such property,
namely
$$(\gamma W)_\kappa=\gamma W+\kappa\partial_A\gamma\gamma^{-1},
\quad W\in{\sans W}_1.\eqno(2.3)$$
The deformation induces an action of $\Gau_0(\Sigma^\circ,L)$
on the functionals $\lambda_X+a$, $X\in{\sans W}_0$,
$a\in{\Bbb C}$:
$$(\gamma(\lambda_X+a))_\kappa(W)=\lambda_X((\gamma^{-1}W)_\kappa)+a
=\lambda_{\gamma X}(W)+a-\kappa\langle X,\gamma^{-1}\partial_A\gamma
\rangle,\quad W\in{\sans W}_1.\eqno(2.4)$$
By combining $(2.2)$ and $(2.4)$, one verifies that the deformed action thus
defined is Poisson. At the infinitesimal level, $(2.3)$ and $(2.4)$ become
$$(\delta_\xi W)_\kappa=\kappa\partial_A\xi+\delta_\xi W,\eqno(2.5)$$
$$(\delta_\xi(\lambda_X+a))_\kappa(W)=-\lambda_X((\delta_\xi W)_\kappa)=
\lambda_{[\xi,X]}(W)+\kappa\langle\xi,\partial_A X\rangle,\eqno(2.6)$$
where $\xi\in\Lie\Gau_0(\Sigma^\circ,L)$ (cf. eq. $(1.15)$).
{}From $(2.2)$ and $(2.6)$, one has
$$(\delta_\xi(\lambda_X+a))_\kappa=\{J_\xi,\lambda_X+a\}_\kappa,\eqno(2.7a)$$
$$J_\xi(W)=\langle\xi,W\rangle,\quad W\in{\sans W}_1.\eqno(2.7b)$$
{}From here, it appears that the deformed action of $\Gau_0(\Sigma^\circ,L)$ on
${\sans W}_1$ is Hamiltonian with respect to the Poisson structure $(2.2)$,
the Hamiltonian functions being the $J_\xi$.
$J_\xi$ can trivially be written as
$$J_\xi(W)=\langle\xi,J(W)\rangle,\quad W\in{\sans W}_1.\eqno(2.8a)$$
$$J(W)=W.\eqno(2.8b)$$
So, the map $W\in{\sans W}_1\rightarrow J(W)\in{\sans W}_1$ can be identified
with the moment map of the Hamiltonian action.

Next consider $\Conf_0(\Sigma^\circ)$. For any
$f\in\Conf_0(\Sigma^\circ)$, one has that
$\chi(f^*X,f^*Y)=\chi(X,Y)-\kappa\langle [X,Y],A-f^{-1*}A\rangle$, where
for $f\in\Conf_0(\Sigma^\circ)$, $f^*A=\partial L_fL_f{}^{-1}+\partial f
\Ad L_fA\circ f$. Because of the non invariance of $\chi$, the action of
$\Conf_0(\Sigma^\circ)$ on ${\sans W}_1$, defined by $(1.16)$, is not
Poisson. However, in this case too, there is a deformation of the action
enjoying this property. Set
$$(f^*W)_\kappa=\kappa(f^*A-A)+f^*W, \quad W\in{\sans W}_1.\eqno(2.9)$$
The deformation induces an action of $\Conf_0(\Sigma^\circ)$
on the functionals $\lambda_X+a$, $X\in{\sans W}_0$, $a\in{\Bbb C}$ given by
$$(f^*(\lambda_X+a))_\kappa(W)=\lambda_X((f^{-1*}W)_\kappa)+a
=\lambda_{f^*X}(W)+a+\kappa\langle X,f^{-1*}A-A\rangle,
\quad W\in{\sans W}_1.\eqno(2.10)$$
{}From $(2.2)$ and $(2.10)$, it follows that the action $(2.9)$ is Poisson.
At the infinitesimal level, $(2.9)$ and $(2.10)$ become
$$(\theta_vW)_\kappa=\kappa\partial_AL_v+\theta_vW,\eqno(2.11)$$
$$(\theta_v(\lambda_X+a))_\kappa(W)=-\lambda_X((\theta_vW)_\kappa)=
\lambda_{\theta_vX}(W)-\kappa\langle X,\partial_A L_v\rangle,\eqno(2.12)$$
where $v\in\Lie\Conf_0(\Sigma^\circ)$ (cf. eq. $(1.17)$). Now, it can
be verified that
$$(\theta_v(\lambda_X+a))_\kappa=\{T_v,\lambda_X+a\}_\kappa,\eqno(2.13a)$$
$$T_v(W)=(1/2\kappa)\langle vW,W\rangle+\langle L_v, W\rangle,
\quad W\in{\sans W}_1.\eqno(2.13b)$$
This shows that the action $(2.11)$ is Hamiltonian, the Hamiltonian functions
being the $T_v$. $T_v$ can be written as
$$T_v(W)=\langle v,T(W)\rangle,\quad W\in{\sans W}_1\eqno(2.14a)$$
$$T(W)=\tr\big({\cal D}^tW+(1/2\kappa)W^2\big),\quad{\cal D}^t=(1/2)t_{+1}
+\partial t_0-(\partial^2+R)t_{-1}.\eqno(2.14b)$$
So the map $W\in{\sans W}_1\rightarrow T(W)\in{\sans K\sans N}_2$ is the
moment map of the Hamiltonian action.

{}From $(2.2)$, $(2.7b)$ and $(2.13b)$, one gets
$$\{J_\xi,J_\eta\}_\kappa=J_{[\xi,\eta]}+\kappa\chi(\xi,\eta),
\quad \xi,\eta\in\Lie\Gau_0(\Sigma^\circ,L),\eqno(2.15)$$
$$\{T_v,T_w\}_\kappa=T_{[v,w]}+12\kappa\tr(t_0{}^2)\sigma(v,w),
\quad v,w\in\Lie\Conf_0(\Sigma^\circ),\eqno(2.16)$$
$$\{T_v,J_\xi\}_\kappa=J_{\theta_v\xi}+\kappa\chi(L_v,\xi),
\quad\xi\in\Lie\Gau_0(\Sigma^\circ,L),~v\in\Lie\Conf_0(\Sigma^\circ),
\eqno(2.17)$$
where $\sigma(v,w)=-{1\over 12}\langle v,D_1w\rangle$ is the KN $1$--cocycle
and $D_1$ is given in $(1.18)$.

Let us know examine the the above results, interpret them and compare them
with the known literature. $(2.15)$ is a Poisson bracket algebra closely
resembling a Kac--Moody algebra of level $\kappa$. The moment map $J(W)$,
eq. $(2.8b)$, plays here the role of the Kac--Moody current. Similarly,
$(2.16)$ is a Poisson bracket Virasoro algebra of central charge
$12\kappa\tr(t_0{}^2)$. This is the well-known value of the classical
central charge encountered in the theory of classical $W$--algebras
\ref{1,2,3}. The moment map $T(W)$, eq. $(2.14b)$, is the
energy-momentum tensor. In the usual approach \ref{1,3},
the central charge originates
from an improvement term added to the Sugawara energy-momentum tensor
of Kac--Moody theory in order to maintain conformal invariance upon
carrying out the Hamiltonian reduction of the Kac--Moody phase space.
The first and second contributions in expression $(2.13b)$ of $T_v$
correspond more or less to such terms in the present formulation.
Here, however, the improvement term is yielded {\it ab initio} by
the nature of the Drinfeld--Sokolov vector bundle and the action of
the conformal group of $\Sigma^\circ$. The second derivative term
appearing in expression $(2.14b)$ of $T(W)$ has a counterpart in the usual
approach where it is added {\it ad hoc} after the reduction of the phase
space \ref{1,3}.
Here, it is present from the beginning and it strictly necessary to
ensure the correct transformation properties of $T(W)$ under coordinate
changes. From $(2.17)$, the current $J(W)$ transforms as a primary field
of conformal weight $1$ under Poisson bracketting, except for the component
corresponding to the generator $t_{+1}$ of $\goth g$ (see eqs. $(1.18)$ and
$(2.2b)$). This also is familiar in the theory of classical $W$--algebras
\ref{1,3}.
\par\vskip.6cm
\item{3.} {\bf The reduction of the Poisson manifold} $({\sans W}_1,\{\cdot,
\cdot\}_\kappa)$
\vskip.4cm
\par
To obtain the classical $W$--algebras in the above framework, one has to
impose a suitable set of constraints on the Poisson manifold $({\sans W}_1,
\{\cdot,\cdot\}_\kappa)$ to reduce it. The constraints have the form
$$J_\xi\approx 0,\quad \xi\in{\cal X},\eqno(3.1)$$
where $\cal X$ is some subset of $\Lie\Gau_0(\Sigma^\circ,L)$. Such
constraints are essentially of the same form as those used in \ref{1}
once one recalls that in the present formulation the counterpart
of the Kac--Moody current is $A+J(W)$. To implement the reduction
of $({\sans W}_1,\{\cdot,\cdot\}_\kappa)$, one demands that the
constraints are first class. From $(2.15)$, this yields the condition
$$[\xi,\eta]\in{\cal X}~{\rm and}~\chi(\xi,\eta)=0,
\quad \xi,\eta\in{\cal X}.\eqno(3.2)$$
One also requires that the constraint manifold is invariant under the action
of $\Conf_0(\Sigma^\circ)$. From $(2.17)$, this yields the condition
$$\theta_v\xi\in{\cal X}~{\rm and}~\chi(L_v,\xi)=0,
\quad v\in\Lie\Conf_0(\Sigma^\circ),\xi\in{\cal X}.\eqno(3.3)$$
A maximal subspace ${\cal X}$ of $\Lie\Gau_0(\Sigma^\circ,L)$
satisfying $(3.2)-(3.3)$ is obtained as follows. The treatment
given here follows very closely that of \ref{1}. Consider the
$2$-form $\omega\in\bigwedge^2{\goth g}^\vee$ defined by $\omega(x,y)=
\tr(t_{+1}[x,y])$, $x,y\in{\goth g}$. The restriction of such form
to ${\goth g}_{-{1\over 2}}$ is non singular
\footnote{${}^2$}{Here, ${\goth g}_m=\{x|
x\in{\goth g},~\ad t_0x=mx\}$, ${\goth g}_{\leq m}=\bigoplus_{k\leq m}
{\goth g}_k$, etc.. The orthogonal complement of a subspace $\goth v$ of
$\goth g$ with respect to the invariant bilinear form defined by the
trace $\tr$ will be denoted by ${\goth v}^\perp$.}.
By Darboux theorem, there is a direct sum decomposition
${\goth g}_{-{1\over 2}}={\goth p}_{-{1\over 2}}\oplus{\goth q}_{-{1\over 2}}$
into subspaces of ${\goth g}_{-{1\over 2}}$ which are maximally isotropic and
dual to each other with respect to $\omega$. Set
$${\goth x}={\goth g}_{\leq -1}\oplus{\goth p}_{-{1\over 2}},\eqno(3.4)$$
which is a nilpotent subalgebra of $\goth g$. Then, one can show that
$${\cal X}=\{\xi|\xi\in\Lie\Gau_0(\Sigma^\circ,L),~\xi~{\rm valued~in}~
{\goth x}\}.\eqno(3.5)$$
It can be proven that the condition of valuedness in $\goth x$ is
compatible with changes of trivializations of $L$ \ref{13}. Such condition
involves no restriction on the KN content of $\cal X$. In fact, it can be
shown that
$${\cal X}\simeq
{\sans K\sans N}_{1\over 2}\oplus\cdots\oplus
{\sans K\sans N}_{1\over 2}\oplus\bigoplus_{\eta\in\Pi,m\in I_\eta,m\geq 1}
{\sans K\sans N}_m,\eqno(3.6)$$
where there are $\dim{\goth p}_{-{1\over 2}}$ ${\sans K\sans N}_{1\over 2}$
spaces in the right hand side. The explicit for of the isomorphism
will be given elsewhere \ref{13}.

The constraint manifold ${\cal W}_{\rm constr}$ is given in terms of the
orthogonal complement ${\goth x}^\perp$ of $\goth x$
$${\goth x}^\perp={\goth g}_{\leq 0}\oplus\ad t_{+1}{\goth p}_{-{1\over 2}}
\eqno(3.7)$$
and is explicitly given by
$${\cal W}_{\rm constr}=\{W|W\in{\sans W}_1,~W~{\rm valued~in}
{}~{\goth x}^\perp\}.\eqno(3.8)$$
Here too, one can show that the condition of valuedness in ${\goth x}^\perp$
is compatible with changes of trivializations of $L$ and that in fact no
restriction on the KN content of ${\cal W}_{\rm constr}$ results in the sense
that
$${\cal W}_{\rm constr}\simeq
{\sans K\sans N}_{-{1\over 2}}\oplus\cdots\oplus
{\sans K\sans N}_{-{1\over 2}}\oplus\bigoplus_{\eta\in\Pi,m\in I_\eta,m\geq 0}
{\sans K\sans N}_m,\eqno(3.9)$$
where the right hand side contains $\dim{\goth p}_{-{1\over 2}}$
${\sans K\sans N}_{-{1\over 2}}$ spaces.

{}From $(1.15)$ and $(2.5)$, it follows that, for $\xi\in{\cal X}$ and
$W\in{\cal W}_{\rm constr}$, $(\delta_\xi W)_\kappa\in{\cal W}_{\rm constr}$.
Similarly, from $(1.17)$, $(1.18)$ and $(2.11)$, it follows that
for $v\in\Lie\Conf_0(\Sigma^\circ)$ and
$W\in{\cal W}_{\rm constr}$, $(\theta_vW)_\kappa\in{\cal W}_{\rm constr}$.
The gauge symmetry, associated to the first class constraints $(3.1)$,
must be fixed.
It can be shown that for any $W\in{\cal W}_{\rm constr}$,
there exists a unique element $\zeta_W\in{\cal X}$ depending polynomially
on $W$, $R$ and their derivatives such that
$$(\exp\zeta_WW)_\kappa=W_c,\eqno(3.10)$$
where $W_c$ is an element of ${\cal W}_{\rm constr}$ such that
$$\ad t_{-1}W_c=0.\eqno(3.11)$$
By the nilpotency of $\goth x$, $W_c$ depends polynomially on $W$, $R$ and
their derivatives as well.
The uniqueness of $\zeta_W$ ensures further that the map $W\rightarrow W_c$
is gauge invariant, {\it i. e.}
$$(\exp\xi W)_{\kappa c}=W_c,\quad \xi\in{\cal X},~W\in{\cal W}_{\rm constr}.
\eqno(3.12)$$
Here, it is important to realize that the standard
proof of the existence and uniqueness of $\zeta_W$ given in refs. \ref{1,3}
cannot be straightforwardly generalized to the present framework, for due
account of global issues is not taken. In spite of this, the result remains
true.

The above suggests the following gauge fixing condition
$$W=W_c,\quad W\in{\cal W}_{\rm red}, \eqno(3.13)$$
defining the reduced manifold ${\cal W}_{\rm red}$.
${\cal W}_{\rm red}$ can be characterized in terms of a set of second
class constraints. Let
$${\cal X}'=\{\xi|\xi\in\Lie\Gau_0(\Sigma^\circ,L),~\xi~{\rm valued~in}~
(\ker\ad t_{-1})^\perp\}.\eqno(3.14)$$
The KN content of ${\cal X}'$ is expressed by the isomorphism
$${\cal X}'\simeq\bigoplus_{\eta\in\Pi,m\in I_\eta,m\geq -j_\eta+1}
{\sans K\sans N}_m.\eqno(3.15)$$
${\cal W}_{\rm red}$ is defined by the second class constraints
$$J_\xi\approx 0,\quad \xi\in{\cal X}'.\eqno(3.16)$$
So, ${\cal W}_{\rm red}$ is given by
$${\cal W}_{\rm red}=\{W|W\in{\sans W}_1,~W~{\rm valued
{}~in}~\ker\ad t_{-1}\}.\eqno(3.17)$$
The KN content of ${\cal W}_{\rm red}$ is expressed by the isomorphism
$${\cal W}_{\rm red}\simeq\bigoplus_{\eta\in\Pi}
{\sans K\sans N}_{j_\eta+1},\eqno(3.18)$$
a relation which could be deduced also from $(3.11)$ and $(3.13)$.
It is readily verified that $(3.3)$ holds with $\cal X$ replaced
by ${\cal X}'$, showing that the reduced manifold is invariant under
$\Conf_0(\Sigma^\circ)$.
${\cal W}_{\rm red}$ equipped with the Dirac brackets $\{\cdot,
\cdot\}_\kappa^*$ defines the reduced Poisson
manifold $({\cal W}_{\rm red},\{\cdot,\cdot\}_\kappa^*)$,
whose properties are the topic of the next section.
\par\vskip.6cm
\item{4.} {\bf The Poisson manifold} $({\cal W}_{\rm red},\{\cdot,
\cdot\}_\kappa^*)$
\vskip.4cm
\par
The task now facing one is the computation of the Dirac brackets $\{\cdot,
\cdot\}_\kappa^*$. Consider the dual space ${\cal W}_{\rm red}^\vee$ of
${\cal W}_{\rm red}$. This can be characterized as
$${\cal W}_{\rm red}^\vee
\simeq\bigoplus_{\eta\in\Pi}{\sans K\sans N}_{-j_\eta}.\eqno(4.1)$$
The dual pairing of ${\cal W}_{\rm red}$ and ${\cal W}_{\rm red}^\vee$
is defined as follows. Let $\nu=(\nu_\eta)_{\eta\in\Pi}$, $\nu_\eta\in
{\sans K\sans N}_{-j_\eta}$ be an element ${\cal W}_{\rm red}^\vee$ and
$\omega=(\omega_\eta)_{\eta\in\Pi}$, $\omega_\eta\in
{\sans K\sans N}_{j_\eta+1}$ be one of ${\cal W}_{\rm red}$. Then
$$\langle\nu,\omega\rangle=\sum_{\eta\in\Pi}N_\eta
\langle\nu_\eta,\omega_{\bar\eta}\rangle\eqno(4.2)$$
(cf. eq. $(1.8)$). The Dirac brackets $\{\cdot,\cdot\}_\kappa^*$ are
completely defined by those of the linear functionals
$$l_\nu(\omega)=\langle\nu,\omega\rangle,\quad\omega\in{\cal W}_{\rm red},
\eqno(4.3)$$
for $\nu\in{\cal W}_{\rm red}^\vee$. The calculation of the Dirac brackets
of the $l_\nu$ involves the choice of a basis of ${\cal X}'$. Luckily,
the explicit expression of the basis elements is not necessary to
carry out the calculation. The result obtained is
$$\{l_\mu,l_\nu\}_\kappa^*(\omega)=\langle[E_{\mu,\kappa^{-1}\omega},
E_{\nu,0}],Q_\omega\rangle+\kappa\chi(E_{\mu,0},E_{\nu,0}),
\quad \mu,\nu\in{\cal W}_{\rm red}^\vee,~\omega\in{\cal W}_{\rm red},
\eqno(4.4a)$$
$$E_{\nu,\omega}=\Big[1+L\ad t_{-1}(\partial_A-\ad Q_\omega)
\Big]^{2j_{\max}+1}P_\nu,
\quad P_\nu=\sum_{\eta\in\Pi}\nu_\eta t_{\eta,j_\eta},
\quad Q_\omega=\sum_{\eta\in\Pi}\omega_\eta t_{\eta,-j_\eta},
\eqno(4.4b)$$
where $L$ is the formal inverse of ${1\over 2}\ad t_{-1}\ad t_{+1}$
extended by $0$ on $\ker\ad t_{+1}$ and $j_{\max}=\max\{j_\eta|\eta\in\Pi\}$.
It can be shown that $E_{\nu,\omega}\in{\sans W}_0$ and $Q_\omega\in
{\sans W}_1$, so that the above expression of the Dirac brackets is globally
defined \ref{13}. The first term in the right hand side of $(4.4a)$ is a
differential polynomial in $\mu$, $\nu$ and $\omega$ and is computable
in principle using $(4.4b)$. The second term, proportional to $\kappa$, is
the anomaly. It can be calculated explicitly. The result is
$$\chi(E_{\mu,0},E_{\nu,0})=\sum_{\eta\in\Pi}N_\eta
\bigg[\prod_{m\in I_\eta, m\geq-j_\eta+1}{2\over C^{-1}{}_{j_\eta,m}}\bigg]
\langle\mu_\eta,D_{j_\eta}\nu_{\bar\eta}\rangle,\eqno(4.5a)$$
$$\eqalignno{D_0=&\hskip3pt\partial,&\cr
D_{1\over2}=&\hskip3pt\partial^2+(1/2)R,&\cr
D_1=&\hskip3pt\partial^3+2R\partial+(\partial R),&\cr
D_{3\over 2}=&\hskip3pt\partial^4+5R\partial^2+5(\partial R)\partial
+(3/2)\big(\partial^2R+(3/2)R^2\big),&\cr
D_2=&\hskip3pt\partial^5+10 R\partial^3+15(\partial R)\partial^2+
[9(\partial^2R)+16R^2]\partial+2[(\partial^3R)+8R(\partial R)],&\cr
&{\rm etc.}.&(4.5b)\cr}$$
The $D_j$ are the well-known Bol operators \ref{14}.

There are other relevant Dirac brackets. Consider the
energy-momentum tensor $T$. For any $v\in\Lie\Conf_0(\Sigma^\circ)$, the
restriction of $T_v$ to ${\cal W}_{\rm red}$ is given by
$$T_v(\omega)=2^{-{1\over 2}}\tr(t_0{}^2)\langle v,\omega_o\rangle+(1/2\kappa)
\sum_{\eta\in\Pi,j_\eta=0}N_\eta\langle v,\omega_\eta{}^{\otimes 2}\rangle
\eqno(4.6)$$
(see below eq. $(1.7b)$). From $(4.4a)$ and $(4.6)$, one has
$$\{T_v,T_w\}_\kappa^*=T_{[v,w]}+12\kappa\tr(t_0{}^2)\sigma(v,w),
\quad v,w\in\Lie\Conf_0(\Sigma^\circ),\eqno(4.7)$$
which is to be compared with $(2.16)$. One further finds that
$$\{T_v,l_\mu\}^*_\kappa=l_{\theta_v\mu}+\kappa\chi(L_v,E_{\mu,0}),
\quad v\in\Lie\Conf_0(\Sigma^\circ),~\mu\in{\cal W}_{\rm red}^\vee,
\eqno(4.8)$$
where $\theta_v\mu_\eta$ is given by $(1.5)$ with $p=\mu_\eta$ and
$j=-j_\eta$.

Let us discuss briefly the results just obtained. $(4.4)$ defines a Dirac
bracket $W$--algebra in the so-called lowest weight gauge. In fact,
analogous expression have been worked out in the literature following
analogous techniques (see f. i. ref. \ref{3} for a review).
The $W$--algebra proper is obtained by letting $\mu_\eta$ and $\nu_\eta$
in $(4.4a)$ be elements of the KN basis of ${\sans K\sans N}_{-j_\eta}$.
The form of the anomaly was first found in \ref{6} in a different
approach where however the deep relation with the theory of $A_1$ embeddings
into simple Lie algebras was not apparent. From $(4.6)$ and $(4.7)$, it
follows that the $T_v$ form a Dirac bracket Virasoro algebra of classical
central charge $12\kappa\tr(t_0{}^2)$. From $(4.8)$, it also appears that
the functions $l_\mu$ with $\mu_o=0$ are primary with respect to the Virasoro
algebra. All the above properties have a counterpart in the standard algebraic
formulation to $W$--algebras \ref{1,3}.

The present paper provides a synthesis of the algebraic and geometrical
approaches to $W$--algebras and shows them in a completely new light.
As a possible application, one may consider a special class of flat forms of
the Drinfeld--Sokolov vector bundle $L$, the ones corresponding to
equivalence classes of elements $W\in{\cal W}_{\rm constr}$ modulo
$G_{\goth x}$ gauge transformations, where $G_{\goth x}$ is the subgroup
of $G$ corresponding to the Lie subalgebra $\goth x$. Such flat forms
may be viewed as functions on ${\cal W}_{\rm red}$. It would be interesting
to compute the Dirac brackets $\{\imath(l_1),\imath(l_2)\}_\kappa^*$ where
$\imath$ is the characteristic homomorphism of $\pi_1(\Sigma^\circ)$
associated to a flat form of $L$ and, for fixed $l\in\pi_1(\Sigma^\circ)$,
$\imath(l)$ is viewed as a function on ${\cal W}_{\rm red}$. This would probe
world sheet topological effects of $W$--algebras. It also would
be interesting to develop a BRST formalism and study quantization.
Lastly, it is also important to analyze the relation between the above
approach to $W$--algebras and the formulation of Toda theory on Riemann
surfaces of ref. \ref{15}, where the basic holomorphic vector bundle
can be shown to be holomorphically equivalent on $\Sigma^\circ$
to the Drinfeld--Sokolov bundle $L$
\footnote{${}^3$}{I thank E. Aldrovandi for pointing this out to me}.
\vskip.4cm
\par\noindent
{\bf Acknowledgements.} I wish to voice my gratitude to
E. Aldrovandi and expecially R. Stora for helpful discussions.
\vskip.6cm
\centerline{\bf REFERENCES}
\def\ref#1{\lbrack #1\rbrack}

\def\PL#1{Phys.~Lett.~{\bf #1}}

\def\PR#1{Phys.~Rev.~{\bf #1}}

\def\LMP#1{Lett.~Math.~Phys.~{\bf #1}}

\def\IJMP#1{Int.~J.~Mod.~Phys.~{\bf #1}}
\def\PREP#1{Phys.~Rep.~{\bf #1}}

\vskip.4cm
\par\noindent

\item{\ref{1}}
L. Feh\'er, L. O'Raifeartaigh, P. Ruelle, I. Tsutsui and A. Wipf, \PR{222}
no. 1 (1992) 1 and references therein.

\item{\ref{2}}
P. Bouwknegt and K. Schoutens, \PREP{223} (1993) 183 and references therein.

\item{\ref{3}} T. Tjin, {\it Finite and Infinite W--algebras and their
Applications}, Doctoral Thesis, LANL hep-th/9308146 and references therein.

\item{\ref{4}} I. M. Krichever and S. P. Novikov, Funktz. Analiz.
Prilozhen. {\bf 21} n. 2 (1987) 46.

\item{\ref{5}} I. M. Krichever and S. P. Novikov, Funktz. Analiz.
Prilozhen. {\bf 21} n. 4 (1987) 47.

\item{\ref{6}} M. Matone, {\it Conformal Field Theories in Higher Genus},
Doctoral Thesis, SISSA--ISAS, Trieste, Italy, and references therein.

\item{\ref{7}} L. Bonora, M. Martellini, M. Rinaldi and J. Russo,
\PL{B206} (1988) 444.

\item{\ref{8}} P. Bryant, \LMP{19} (1990) 97.

\item{\ref{9}}
L. Bonora, M. Rinaldi, J. Russo and K. Wu, \PL{B208} (1988) 440.

\item{\ref{10}}
J. De Boer and J. Goeree, \PL{B274} (1992) 289 and
preprint LANL hep-th/9206098.

\item{\ref{11}} R. Zucchini, LANL hep-th/9307015.

\item{\ref{12}}
V. G. Drinfeld and V. V. Sokolov, J. Sov. Math. {\bf 30} (1985) 1975.

\item{\ref{13}} R. Zucchini, in preparation.

\item{\ref{14}}
F. Gieres, \IJMP{A8} (1993) 1.

\item{\ref{15}}
E. Aldrovandi and L. Bonora, LANL hep-th/9303064.
\bye